\documentclass{emulateapj}

\usepackage[colorlinks=true,linkcolor=blue,citecolor=blue]{hyperref}

\slugcomment{Draft \today}

\shorttitle{The Morphology  and Dynamics of Jet-Driven Supernova Remnants}

\shortauthors{Gonz\'alez-Casanova et al.}

\begin{document}

\author{Diego F. Gonz\'alez-Casanova\altaffilmark{1}, Fabio De Colle\altaffilmark{1}, Enrico Ramirez-Ruiz\altaffilmark{2}, 
and Laura A. Lopez\altaffilmark{3,4,5}}
\altaffiltext{1}{Instituto de Ciencias Nucleares, Universidad Nacional Aut{\'o}noma de M{\'e}xico, A. P. 70-543 04510 D. F. Mexico}
\altaffiltext{2}{Department of Astronomy and Astrophysics, University of California, Santa Cruz, CA 95064, USA}
\altaffiltext{3}{MIT-Kavli Institute for Astrophysics and Space Research, 77 Massachusetts Avenue, 37-664H, Cambridge, MA 02139, USA}
\altaffiltext{4}{NASA Einstein Fellow}
\altaffiltext{5}{MIT Pappalardo Fellow in Physics}

\title{The Morphology  and Dynamics of Jet-Driven Supernova Remnants: the Case of W49B}

 \begin{abstract} 
The circumstellar medium (CSM) of a massive star is modified by its winds before a supernova (SN) explosion occurs, and thus the evolution of the resulting supernova remnant (SNR) is influenced by both the geometry of the explosion as well as the complex structure of the CSM. Motivated by recent work suggesting the SNR W49B was  a jet-driven SN expanding  in a complex CSM, we explore how the dynamics and  the metal distributions in  a jet-driven  explosion  are modified by the  interaction with  the surrounding  environment. In particular, we perform  hydrodynamical calculations to study the dynamics and explosive nucleosynthesis of a jet-driven SN triggered by the collapse of a  25 $M_{\sun}$ Wolf-Rayet star and its subsequent interaction with the CSM up to several hundred years following the explosion. We find that although the CSM  has small-scale effects on the structure of the SNR, the overall morphology and abundance patterns are reflective of  the initial  asymmetry of the SN explosion. Thus, we predict that jet-driven SNRs, such  as W49B,  should be identifiable based on morphology and abundance patterns at ages up to several hundred years, even if they expand into a complex CSM environment.  
\end{abstract}


\section{Introduction}
The morphology and dynamics of young supernova remnants (SNRs) depend on the geometry of the explosions as well as the distribution of the circumstellar medium (CSM; see review by \citealt{vink12}). SNRs from massive progenitors are luminous from the interaction of the ejecta with nearby gas that was expelled by the progenitor system \citep{weaver1977}, probably during a red supergiant stage. Beyond the relic red giant wind material, the CSM is likely to have been highly disturbed by previous evolutionary stages \citep{garcia1996a,garcia1996b,ramirez-ruiz2005} as well as by density discontinuities or gradients in the interstellar medium (ISM; \citealt{dpj1996}), such as molecular cloud  edges. Consequently, the relative role of the explosion and the environment in shaping SNRs remains an outstanding question, and it has even been cited as the biggest challenge in modern SNR research \citep{canizares04}.

The difficult task of unraveling the origin of asymmetries in SNRs is best illustrated by W49B (G43.3$-$0.2), the most  luminous SNR in our Galaxy in X-rays ($L_{\rm X} \sim 10^{38}$ erg s$^{-1}$; \citealt{immler2005}) and in $\gamma$-rays \citep{abdo2010}. X-ray imaging of W49B has shown an unusual morphology, with a highly elliptical shape comprised of an iron-rich bar with two plumes at its edges  \citep{hwang2000,miceli2006,lopez2009}. The integrated X-ray spectrum exhibits  strong emission lines from several metals (e.g., Si, S, Ar, Ca, and Fe: \citealt{hwang2000,keohane2007}) with supersolar abundances, indicating an ejecta origin. These metals appear segregated, with the Fe absent in the east, while the silicon and other intermediate-mass elements are distributed more homogeneously \citep{lopez2009,lopez2013b}.  Two scenarios have been proposed to explain the anomalous morphology of W49B: a bipolar/jet-driven core-collapse (CC) SN \citep{keohane2007,lopez2013b} or a spherically symmetric SN that expanded  into an inhomogeneous ISM \citep{zhou2011}. 

Observational evidence shows that W49B has a complex environment: near-infrared imaging revealed bright [Fe {\sc ii}] rings thought to be wind material from a massive star progenitor \citep{keohane2007}. Additionally, $^{13}$CO and warm, shocked H$_{2}$ gas have been detected surrounding the SNR \citep{simon2001,keohane2007}. However, the abundance ratios of the ejecta from across W49B are similar to the predictions for a jet-driven explosion of a 25 $M_{\sun}$ Wolf-Rayet (WR) progenitor  \citep{maeda2003} and are inconsistent with values expected in symmetric CC SNe \citep{lopez2013b}. Moreover, the distinct morphology of Fe is likely due to anisotropic ejection of heavy metals, since Fe {\sc xxv} line emission is missing from the eastern part of the SNR, even though the temperatures there are sufficient to produce that line emission.  

These observations show that the X-ray line emission morphology of young SNRs like W49B can be altered not only by  the  structure  of the ejecta but also by the spatial distribution of the elements synthesized within the SN  explosion. Numerical models of jet-driven SN explosions \citep[e.g.][]{Khokhlov99, Couch2009} show that the greater kinetic energy at the poles of the exploding star can lead to efficient synthesis of nickel there, while lower-$Z$ elements are ejected more isotropically \citep[e.g.][]{mazzali2005,nomoto2006}. The issue of how this  bipolar morphology is modified by the expansion into the external medium \citep{lopez2009} and its implications for W49B remains an unresolved problem. 

In this {\it Letter}, we aim to determine the degree to which the dynamics and abundance structures of a jet-driven SNR are modified by the interaction with the external medium. As the CSM of a massive star is modified by their winds, we consider shocked wind bubbles for the surrounding media. Given that a complete simulation of a jet-driven SNR's evolution is computationally expensive, we perform two separate hydrodynamical calculations, each equipped to describe the behavior of the bipolar outflow at two different epochs. In Section~\ref{sec:num}, we present a brief description of the numerical methods. In Section~\ref{sec:jet}, we use a simulation with a general equation of state and a reaction network to determine the nucleosynthesis that accompanies a jet propagating through the massive star. In Section~\ref{sec:snr}, we employ the resulting models (i.e., the density, velocity, and compositional structure of the ejecta) as the initial conditions in the calculation of the subsequent hydrodynamical expansion of the SNR and the associated thermal X-ray emission produced by the hot shocked gas. Finally, in Section~\ref{sec:dis} we summarize our results and present our conclusions.

\section{Numerical Methods}\label{sec:num}

In order to study the dynamics and explosive nucleosynthesis of a jet-driven SN, we have performed a set of two-dimensional cylindrically symmetric 
simulations. We use  the adaptive mesh refinement code \emph{Mezcal} \citep{decolle2006,decolle2008,decolle2012a,decolle2012b},  which incorporates  a nuclear network module and different equations of state (EOS). The {\it Helmholtz} EOS  \citep{timmes2000b}, which includes radiation, ion pressure, and electron-positron degeneracy pressures, is used in Section~\ref{sec:jet} when calculating the propagation of the jet through the star, while a simple ideal gas EOS  is used in Section~\ref{sec:snr} when computing the long-term evolution of the SNR.

The nucleosynthesis is calculated by coupling the hydrodynamics equations to an inexpensive, seven element {\it $\alpha$-chain} nuclear network \citep{timmes2000a}, comprising the evolution of the $^4$He, $^{12}$C, $^{16}$O, $^{20}$Ne, $^{24}$Mg, $^{28}$Si, 
$^{56}$Ni mass densities and energy losses due to neutrino cooling. In the nuclear network, the $^{28}$Si and $^{56}$Ni mass fractions 
are representative of the silicon and iron mass fraction groups, respectively. A detailed description of the implementation of the nuclear burning module in the \emph{Mezcal} code is presented in \citet{holcomb2013}.

\section{A Jet-Driven Supernova Model for W49B} \label{sec:jet}

\begin{figure}[]
  \centering\includegraphics[height=\linewidth]{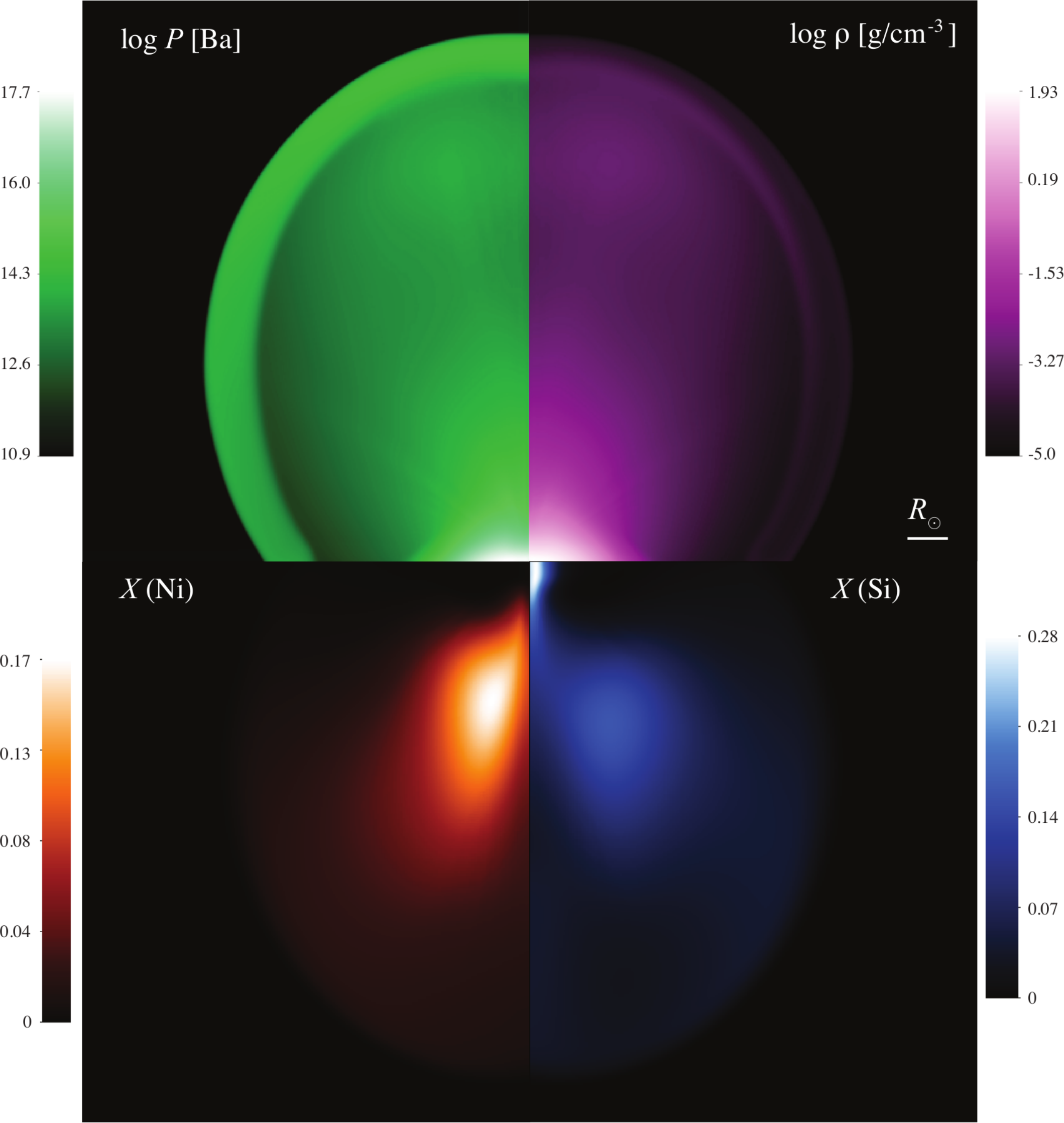}
  \caption{Jet-driven SN explosion at the onset of the homologous phase, which is achieved $100$~seconds after the bipolar ejecta has broken free from the $R_\ast = 0.43 R_\odot$  stellar progenitor. The pressure $P$, mass  density $\rho$ and the silicon $X$(Si)  and nickel $X$(Ni)  mass fractions are indicated in each frame with corresponding size scales in units of $R_\odot$. The  jet propagation inside  the WR   progenitor  is calculated using an adaptive cylindrical grid of physical size $l_{\rm r}=l_{\rm z}= 4.5\times10^{10}$~cm, with $10^2\times10^2$ cells on the coarsest grid and 7 levels of refinement, which corresponds to a maximum resolution of $7\times10^6$~cm.} \label{fig1}
\end{figure}	

\begin{figure}[]
  \centering\includegraphics[height=0.9\linewidth]{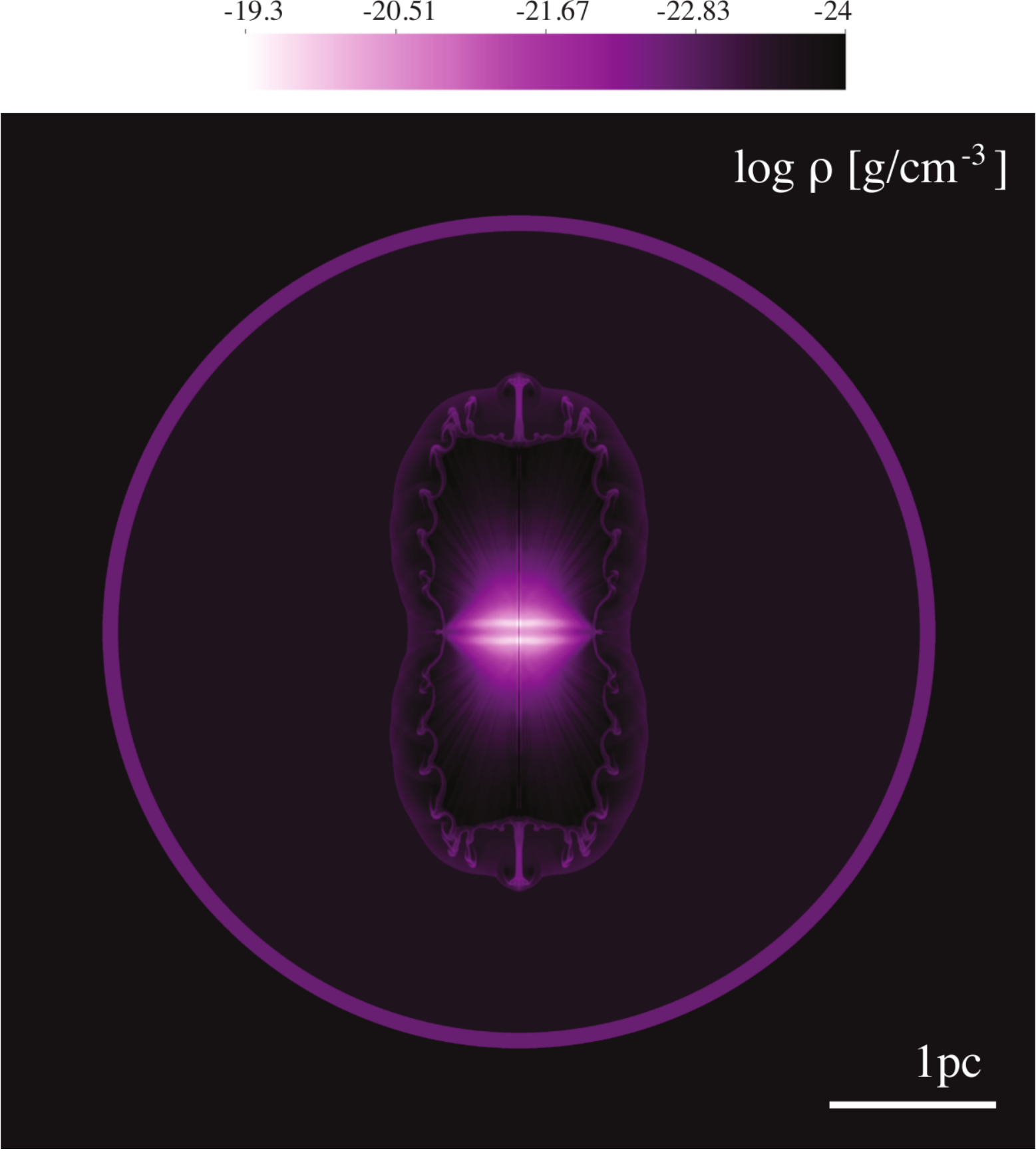}
  \caption{The evolution of a jet-driven SN remnant inside the wind bubble structure expected around a 25 $M_\odot$ massive star for  the case of an ISM pressure typical of the high-density molecular  region surrounding W49B.  The dense RSG  shell, with a mass of about 6.5 $M_\sun$, is  located at 2.5~pc.   The region outside the red supergiant shell is the remains of the bubble from the main-sequence phase while the density in the shocked region is approximately constant. The output of the hydrodynamic simulations at the time of homology (Figure~\ref{fig1}) was mapped onto the grid at small spatial scales. Shown is the logarithm of the mass density  at an evolutionary time of 20 yr after the supernova explosion.  The calculations are performed on a grid with size  $l_{\rm r}=l_{\rm z}= 5\times 10^{19}$~cm,
with $10^2 \times10^2$  cells on the coarsest grid and 7 levels of refinement.}
  \label{fig2}
\end{figure}

Based on the abundances derived from the X-ray spectra of W49B, \cite{lopez2013b} showed the nucleosynthesis was consistent with a bipolar explosion of a 25 $M_{\sun}$ Wolf-Rayet (WR) star. Thus, for our initial conditions, we adopt the {\it E25} pre-supernova stellar model of \cite{heger2000}. The pre-supernova star, initially a 25 $M_{\sun}$ WR star, has a mass  of 5.45 $M_\odot$, a radius of $R_\ast = 0.43 R_\odot= 3 \times 10^{10}$~cm, and an iron  core mass of 1.45 $M_{\sun}$ extending out to a radius $R_{\rm Fe}  = 2.2 \times 10^8$~cm.  A uniform jet with sharp edges and opening angle of $\theta_{\rm j}=30^{\circ}$  is injected from a boundary located at $R_{\rm Fe}$ with a velocity $v_{\rm j} = 0.3c$.  The jet injection time is $\Delta t = 6$~s, which results in a total energy of 6$\times 10^{51}$~erg.

The results of the simulation of the jet-driven SN explosion are presented in Figure~\ref{fig1}, where the mass density $\rho$, pressure $P$, silicon $X$(Si) and nickel $X$(Ni) mass fractions of the expanding collimated ejecta are plotted $10^2$~s after the jet has broken free from the star.  In the direction of the propagation of the jet, the resulting  post-shock temperatures are high (with $T \gtrsim 5 \times 10^9$~K), leading to complete oxygen and silicon burning there \citep{Maeda2002,maeda2003,Maeda2009}. Consequently, the expansion velocity of the newly synthesized $^{56}$Ni is closely aligned with the jet axis. 

As it expands inside the progenitor star, the jet is unable to move the stellar material at a velocity  comparable to its own and  hence is abruptly slowed down. A large fraction of the energy output during this phase is  accumulated into a cocoon surrounding the jet \citep[e.g.,][]{ramirez-ruiz2002}.
Thus, the lateral production of $^{56}$Ni is confined to the deepest stellar layers, and elements ejected in this direction by the cocoon's expansion are primarily the products of hydrostatic nuclear burning 
with some explosive oxygen-burning products (e.g., Si, S). As a result, the distribution of nickel is much more elongated than that of silicon \citep{Maeda2002,maeda2003,Maeda2009}, as can be clearly seen in the abundance distributions shown in Figure \ref{fig1}. 
The jetted outflow, responsible for synthesizing $^{56}$Ni, is observed to carry more energy and inertia than the broader (incomplete silicon and oxygen burning) component,  so that the latter never overtakes it. Therefore, the distribution asymmetry in the abundance is expected to be preserved as the SN material sweeps up the external medium.

\section{The Evolution of a Young Jet-Driven SNR}
\label{sec:snr}

Next, we consider the interaction of the jet with the CSM. The pre-supernova stellar wind depends on the evolutionary stages prior to and during the WR phase. A standard evolutionary track at solar metallicity \citep[e.g.][]{garcia1996a,garcia1996b} is for an O star to evolve through a red supergiant (RSG) phase or luminous blue variable phase with considerable mass loss \citep[e.g.][]{ramirez-ruiz2001}, and then to become a WR star. At low metallicity and for some binary stars, the RSG phase may be absent \citep{izzard2004}. When the fast WR wind starts blowing, it sweeps up the RSG wind material into a dense, cold shell \citep{chevalier2004}.  The radius of the shocked region depends on the duration of the WR phase $t_{\rm wr}$, the ratio of the mass-loss rates $\dot{M}_{\rm wr}/\dot{M}_{\rm rsg}$, and wind velocities  $v_{\rm wr}/v_{\rm rsg}$. Here we assume  $t_{\rm wr}= 3 \times 10^4$ yr, $\dot{M}_{\rm wr}/\dot{M}_{\rm rsg}=10^{-2}$ and $v_{\rm wr}/v_{\rm rsg}=10^2$ \citep{mandm1994}. 

With these reference values, the radius of the wind termination shock is $R_{\rm t}=0.3$ pc and the pressure  (in units of Ba/$k$, where $k$ is the Boltzmann constant)  in the shocked region is $P_{\rm s}/k=2.5 \times 10^{4}$ K cm$^{-3}$ \citep{Chevalier1983}. If the external pressure $P_{0} > P_{\rm s}$ (as expected in molecular clouds where $P_{0}/k = 10^{5}$ K cm$^{-3}$: \citealt{blitz1993}), the expansion of the RSG will be stalled. The density in the shocked region is approximately constant, and beyond the contact discontinuity (at $R_{\rm c}\approx 3R_{\rm t}$: \citealt{Chevalier1983}), a region of shocked (dense) cold RSG material is expected. 

The CSM around a 25 $M_\odot$ massive stellar progenitor appears as in Figure~\ref{fig2}. At solar metallicity, the star has a long-lived RSG phase that dramatically affects the surrounding medium. In particular, a dense RSG shell occurs at roughly  2.5 pc, approximated here for the case of a high ISM pressure typical of molecular clouds. We note that in reality, the RSG shell is subject to hydrodynamical instabilities \citep{garcia1996a,garcia1996b}, and its radius depends sensitively on several parameters, including $t_{\rm wr}$ (which varies with mass and metallicity), $v_{\rm wr}/v_{\rm rsg}$,  $\dot{M}_{\rm wr}/\dot{M}_{\rm rsg}$ and $P_{0}$.

\begin{figure}[]
  \centering\includegraphics[height=0.9\linewidth]{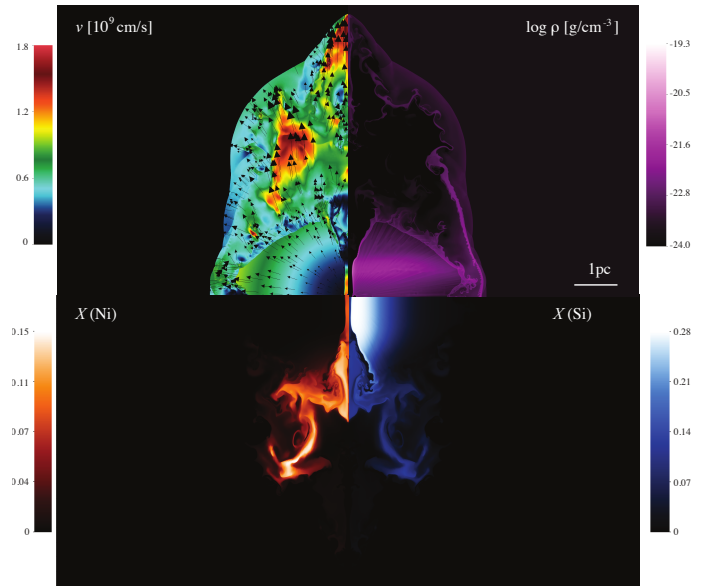}
  \caption{The evolution of a jet-driven SN remnant after the  interaction with the wind bubble structure shown in Figure~\ref{fig2}. Shown  in each frame are the velocity $v$, mass  density $\rho$ and the silicon $X$(Si)  and nickel $X$(Ni)  mass fractions at  $t = 260$ yr  together with the corresponding unit size  scale. The superimposed vectors in the upper left frame show the velocity field of the resulting SN remnant.}
  \label{fig3}
\end{figure}	

Before the SNR  size becomes comparable to the scale length  of the RSG shell, the morphology of the remnant will reflect the geometry of the initial explosion \citep{ramirez-ruiz2010}. This can be seen by comparing the density contours in Figures~\ref{fig1} and \ref{fig2}. 
The  expansion of the bipolar remnant will be  stalled when it reaches the sharp density discontinuity at  $\sim$2.5 pc. The interaction of the ejecta with the  RSG shell  results in the disruption of the dense thin layer  before the jet undergoes substantial lateral expansion. Consequently, the initial asymmetry of the explosion is preserved
(Figure~\ref{fig3}). The RSG shell, clearly apparent in Figure~\ref{fig2}, will be pushed outward at high velocities as a result of the interaction.
Figure~\ref{fig3} also shows the development  of {\it fingers} extending  from the deformed  shell and produced by the Rayleigh-Taylor instability. Despite the large density contrast, the pressure in the remnant remains  fairly uniform. Both the  deformed  thin layer   and the protruding {\it fingers} expand at velocities $\approx$ 500~km s$^{-1}$. These large velocities are produced as a result of  successive shocks  ramming into the deformed thin layer. The strands of material formed as the thin dense  layer is disrupted could be  responsible for  the prominent X-ray line emission filaments  observed in W49B \citep{keohane2007,lopez2013b,peters2013}. 

The accompanying evolution of the metal  distribution is shown in Figure~\ref{fig4}. A broadening of the iron-group filaments is clearly seen  after the interaction with the RSG shell, which is accompanied by the development   of small-scale density fluctuations.
After $t$ = 260 yr and radii of $r \approx$ 3 pc, the iron-group ($X_{\rm Ni}$) and silicon ($X_{\rm Si}$) structures have become highly irregular,  with the iron-group elements 
being significantly more extended  along  the jet axis.  
The remnant also shows a complex velocity structure with fast-moving iron  filaments 
expanding  along  the explosion center, which  are less pronounced in the initially more centrally concentrated silicon-rich material. Thus, although the structure of the CSM plays a prominent role, our results 
demonstrate that  the measurable properties of young SNRs depend sensitively on the initial distribution of the metal-rich ejecta. 

\section{Discussion} \label{sec:dis}

\begin{figure}[t]
  \centering\includegraphics[height=0.55\linewidth]{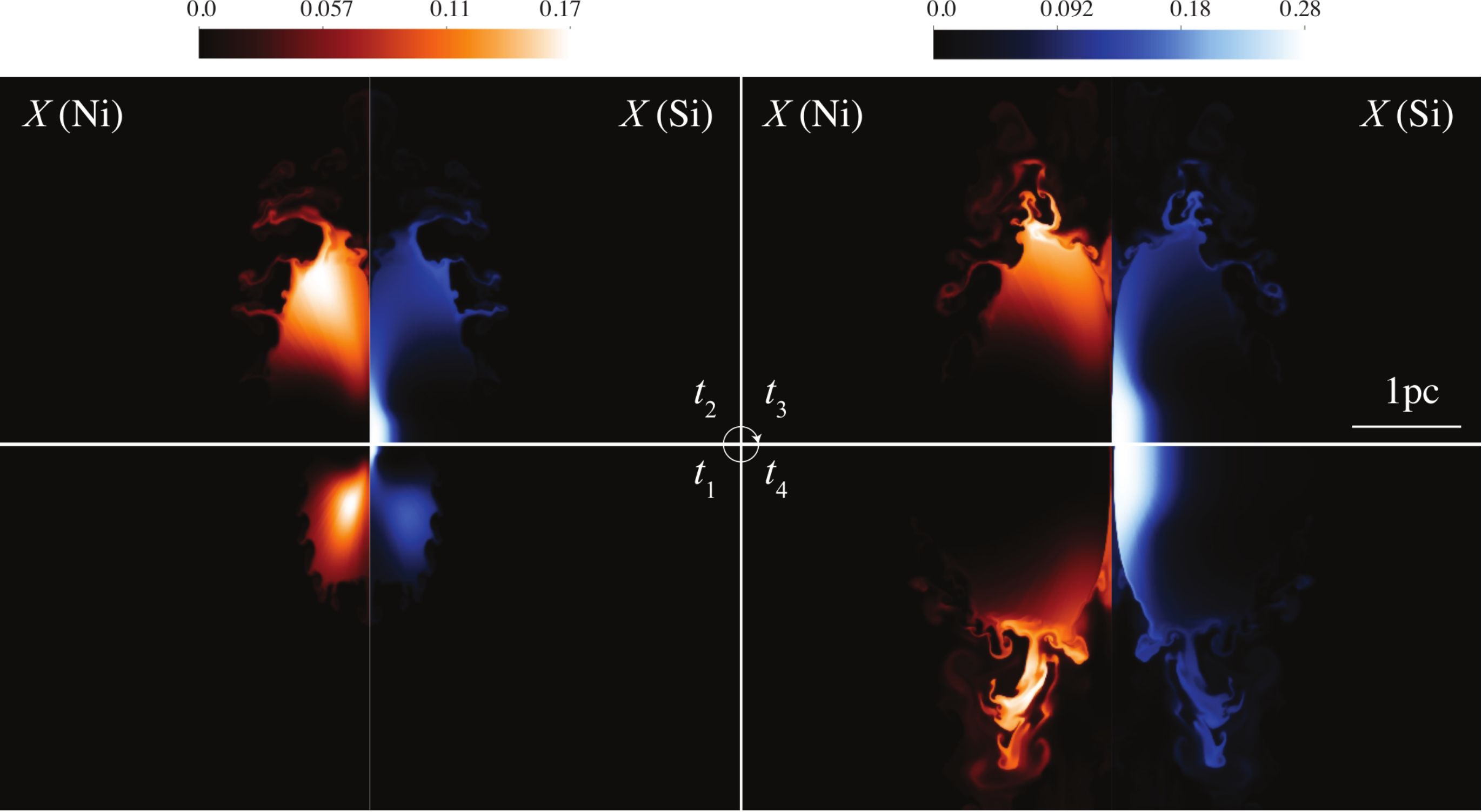}
  \caption{Shown is the evolution of the nickel ($X_{\rm Ni}$) and silicon ($X_{\rm Si}$) mass fractions  at $t$= 60~yr ($t_1$), 100~yr ($t_2$), 140~yr ($t_3$), 260~yr ($t_4$). The individual frames have been successively rotated by $\pi$.  A 1 pc scale bar is shown.}
  \label{fig4}
\end{figure}	

\begin{figure}[t]
  \centering\includegraphics[height=1\linewidth]{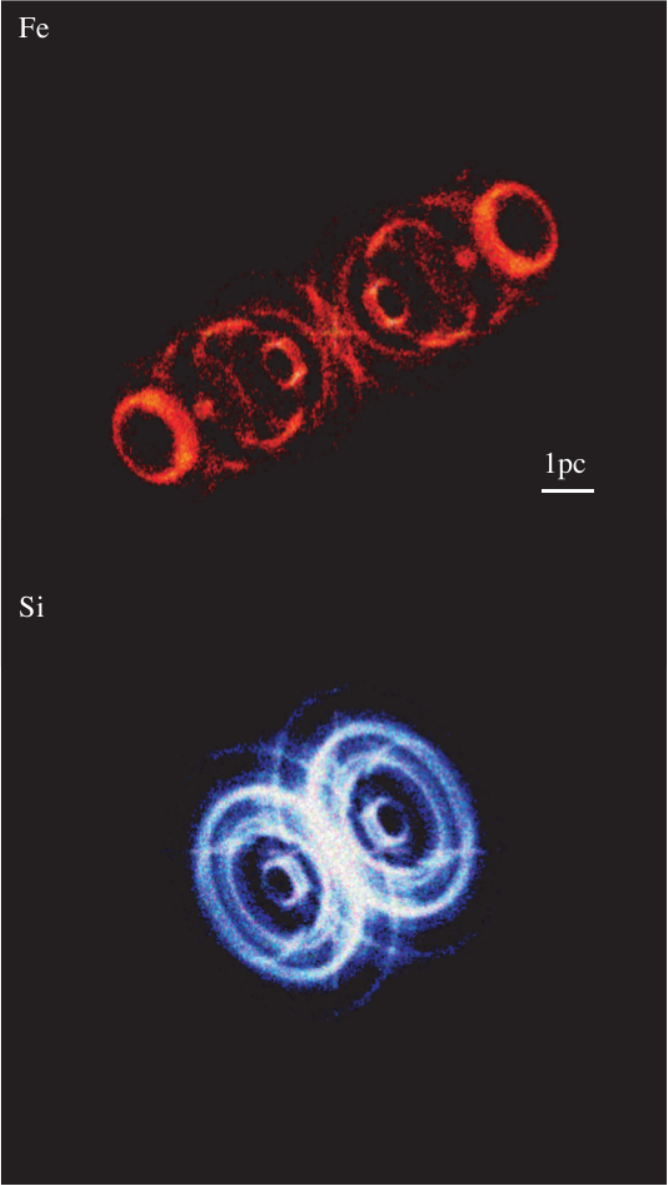}
  \caption{Fe {\sc xxv} and Si   {\sc xiv} X-ray emission maps. The maps are obtained integrating  
             the respective emission coefficients  
              along the line of sight assuming collisional ionization equilibrium \citep{lopez2009}. The images are computed at an evolutionary age  of 700~yr,
              and employ 200 $\times$ 200 pixels. 
              To match the characteristics of the W49B, the results of the simulation have been
              rotated 45$^\circ$ with respect to the plane of the sky (in the direction of the observer), 
              and 120$^\circ$ with respect to the $z$-axis (the initial direction of the jet), before
              the emission being integrating along the line of sight. }
  \label{fig5}
\end{figure}

The results presented here illustrate how the dynamics of young, jet-driven SNRs,
are influenced by  the interaction of the ejecta with the CSM.
By comparing our hydrodynamical simulations to the observational properties of W49B, we can gain a deeper understanding on the nature of bipolar SNR.
A few observables are relatively easy to determine from one-dimensional hydrodynamical models, e.g. expansion rates and thickness of flow structures. However, other comparisons can be more challenging. For example, to predict the distribution of the observed thermal X-ray emission, multidimensional simulations are necessary to follow the detailed distribution and composition of ejecta, their clumping, and the inhomogeneous CSM (e.g., \citealt{badenes03}). 

To date, simulations to reproduce the thermal X-ray morphology of W49B have not been done due to an absence of a commonly accepted dynamical model.  The observed anisotropic distribution of strong X-ray line features from Si and Fe in W49B strongly suggests that heavy-element-enriched material was produced asymmetrically deep within the exploded star \citep{lopez2013b}. The calculations presented here support this idea. Figure~\ref{fig5} shows that even in the presence of a uniform external medium, the thermal X-ray morphology of a jet-driven SNR is dominated by a bright iron jet, rimmed by a plateau of explosive oxygen-burning products, such as silicon and sulfur. The bright X-ray luminosity of W49B may be characteristic of the interaction with the CSM produced by extensive mass loss prior to the explosion (Figure~\ref{fig2}). This CSM  has a mass comparable to that of the supernova material and is not expected to spread out  beyond  a few parsecs in length
because of  the large surrounding pressure. 
 Only after several centuries, the supernova ejecta will break free from this relatively dense CSM (Figure~\ref{fig4}). 

But W49B is also interacting with a molecular cloud in the east \citep{simon2001,keohane2007}. Detailed studies of the X-ray surface brightness along its circumference presented by \citet{lopez2013a}, revealed the stalling of the expansion in the eastern region as the ejecta collides with molecular material.  This interaction is likely responsible for halting the expansion of the jetted ejecta, creating the observed thermal X-ray wing feature from the lateral expansion of shocked material \citep{lopez2013a,lopez2013b}.
As relativistic ions are accelerated  through this  dense molecular material  by highly supersonic shocks, the resultant pion decay  is likely to be responsible for producing  the observed GeV $\gamma$-rays \citep{abdo2010}.
In the future, three-dimensional simulations incorporating an anisotropic external medium will be necessary to test  our two-dimensional results and to thoroughly investigate  the effects of the molecular cloud interaction on the resulting thermal X-ray structures. 

Based on the absence of an X-ray luminous compact object, \cite{lopez2013b} predicted a black hole was formed during the explosion of W49B. If it originated from a jet-driven CC SN, it could have been one with a sub-relativistic collimated  outflow, as opposed to the rare, high Lorentz factor jets in gamma-ray bursts \citep{neil2009}. For example, it could be the common jet-driven SN discussed by \cite{Khokhlov99}, in which case the core collapse continues to form a black hole instead of halting upon producing a neutron star. 

Even for the simplest CSM distribution, we have 
shown that the structure resulting from the expansion of a jet-driven remnant like W49B is considerably
different from  that of a more standard CC SN explosion.
 In particular, the morphology at early stages (after several hundred years) still reflects the asymmetry of the explosion, even in a complex, dense  CSM (Figure~\ref{fig1}). In principle then, one could identify jet-driven SNRs based on the abundance distribution and morphology, although the CSM influences their overall shape somewhat, especially at late times. 

\acknowledgments 
We thank D. Castro, G. Garcia-Segura, A. MacFadyen, R. Margutti, S. Pearson and A. Soderberg  for insightful discussions. 
We acknowledge support from the David and Lucile Packard Foundation, NSF grant AST�0847563 and the DGAPA-PAPIIT-UNAM grant IA101413-2. Support for LAL was provided by NASA through the Einstein Fellowship Program, grant PF1�120085, and the MIT Pappalardo Fellowship in Physics.

\end{document}